\documentclass[conference,10pt]{IEEEtran}
\usepackage[utf8]{inputenc}
\usepackage{url}
\usepackage{graphicx}
\usepackage{multirow}
\usepackage{siunitx}
\usepackage[acronym,nohypertypes=acronym]{glossaries}
\usepackage{balance}

\usepackage[T1]{fontenc} 
\usepackage{amsmath}
\usepackage[cmintegrals]{newtxmath}
\usepackage{bm} 

  \usepackage[caption=false, font=footnotesize]{subfig}

\usepackage[capitalise]{cleveref}

\setacronymstyle{long-short}
\newacronym{MCV}{MCV}{mission-critical voice}
\newacronym{NIST}{NIST}{National Institute of Standards and Technology}
\newacronym{QoE}{QoE}{quality of experience}
\newacronym{MtE}{MtE}{mouth-to-ear}
\newacronym{PTT}{PTT}{push to talk}
\newacronym{PSCR}{PSCR}{Public Safety Communications Research Division}
\newacronym{NTIA}{NTIA}{National Telecommunications and Information Administration}
\newacronym{ITS}{ITS}{Institute for Telecommunication Sciences}

\IEEEtriggeratref{20}
\IEEEtriggercmd{\balance}

\begin{document}
\title{Conducting Mission-Critical Voice Experiments with Automated Speech Recognition and Crowdsourcing}

\author{
  \IEEEauthorblockN{%
    Jan Janak\IEEEauthorrefmark{1},
    Kahlil Dozier\IEEEauthorrefmark{1},
    Lauren Berny\IEEEauthorrefmark{2},
    Liang Hu\IEEEauthorrefmark{1},
    Dan Rubenstein\IEEEauthorrefmark{1},
    Charles Jennings\IEEEauthorrefmark{3},
    Henning Schulzrinne\IEEEauthorrefmark{1}%
  }
  \IEEEauthorblockA{%
    \IEEEauthorrefmark{1}Columbia University, New York, USA%
  }
  \IEEEauthorblockA{%
    \IEEEauthorrefmark{2}University of Oregon, Eugene, USA%
  }
  \IEEEauthorblockA{%
    \IEEEauthorrefmark{3}CUNY, New York, USA%
  }
}

\maketitle



\begin{abstract}

Mission-critical voice (MCV) communications systems have been a critical tool for the public safety community for over eight decades. Public safety users expect MCV systems to operate reliably and consistently, particularly in challenging conditions. Because of these expectations, the Public Safety Communications Research (PSCR) Division of the National Institute of Standards and Technology (NIST) has been interested in correlating impairments in MCV communication systems and public safety user quality of experience (QoE). Previous research has studied MCV voice quality and intelligibility in a controlled environment. However, such research has been limited by the challenges inherent in emulating real-world environmental conditions. Additionally, there is the question of the best metric to use to reflect QoE accurately.

This paper describes our efforts to develop the methodology and tools for human-subject experiments with MCV. We illustrate their use in human-subject experiments in emulated real-world environments. The tools include a testbed for emulating real-world MCV systems and an automated speech recognition (ASR) robot approximating human subjects in transcription tasks. We evaluate QoE through a Levenshtein Distance-based metric, arguing it is a suitable proxy for measuring comprehension and the QoE. We conducted human-subject studies with Amazon MTurk volunteers to understand the influence of selected system parameters and impairments on human subject performance and end-user QoE. We also compare the performance of several ASR system configurations with human-subject performance. We find that humans generally perform better than ASR in accuracy-related MCV tasks and that the codec significantly influences the end-user QoE and ASR performance.

\end{abstract}

\glsresetall

%
%
%
%

%

\section{Introduction}
\label{sec:introduction}

Mission-critical voice (MCV) communications systems, or land-mobile radio (LMR), are a critical tool for the public safety community, including firefighters, police, emergency services, and disaster responders. The community has been relying on LMR systems for more than eight decades. Yet, despite considerable technological differences, LMR systems generally rely on half-duplex two-way push-to-talk (PTT) radios with individual and group talk capabilities. ``Public safety users have an expectation that these systems will function for them reliably and consistently, particularly when performing difficult job tasks in challenging conditions and stressful situations. Unlike commercial cellular systems, LMR systems have been optimized to operate with a high probability of successful transmission and reception for the users.''~\cite{funding}

The Public Safety Communications Research (PSCR) Division of the National Institute of Standards and Technology (NIST) has been interested in correlating impairments in the MCV communication system and public safety user quality of experience (QoE). Previous research has studied MCV voice quality and intelligibility in a controlled environment. However, such environments typically cannot emulate real-world operating environments, i.e., deployed MCV communication systems and channel conditions. NIST PSCR seeks to develop the tools, methodology, and datasets to understand the influence of a broader set of MCV system impairments (e.g., channel access time) on the public safety user QoE.

The traditional method to evaluate end-user QoE relies on human-subject studies performed in a controlled environment emulating the real-world environment~\cite{dumas1999practical}. In the study, human-subject volunteers interact with the communication system under the supervision of an experimenter. The study generates data that are then evaluated to answer research questions and prove or disprove hypotheses about the technology's suitability for a particular communication model or scenario. The choice of which metric to use to characterize end-user QoE in the first place is also a question to be considered. The ``quality" of a user's experience is an inherently subjective matter, and the choice of method to characterize it is ultimately the choice of the designer of the experiment.

We present our efforts to develop the methodology and tools for human-subject experiments with MCV and illustrate their use in human-subject experiments as well as choose a metric to measure end-user QoE. The tools include a configurable testbed for emulating real-world MCV communications systems and an automatic speech recognition (ASR) robot capable of approximating human-subject performance in listening transcription experiments. The metric we use for QoE evaluation is fundamentally based on the Levenshtein Distance~\cite{levenshtein1966binary} metric, which is a way of quantifying the ``edit distance" between two strings. We conduct two human-subject studies to understand the influence of system parameters, such as the selected voice codec, and impairments, such as the bit error rate (BER), on human subject performance and the end-user QoE. We perform listening (transcription) experiments with local human subjects, remote Amazon MTurk human subjects, and an ASR robot emulating human behavior.

We find that human subject performance consistently declines as the bit error burst size increases under the P.25 Phase 2 codec but not the AMR wideband codec. Humans perform consistently better than all tested ASR robot configurations, and the difference is especially apparent for large bit-error bursts.



The rest of the paper is organized as follows. In \cref{sec:motivation}, we provide a brief motivation for the research. \Cref{sec:purpose} discusses the study's overall purpose. \Cref{sec:experimental-setup} describes the experimental setup and procedures and a brief overview of the Levenshtein distance algorithm used in the experiments. This is followed by a short discussion of the analytical techniques and methodology in \cref{sec:analytical-procedures}. \Cref{sec:results,sec:discussion} present and discuss the results obtained in our experiments. We summarize related work in \cref{sec:related-work}. \cref{sec:conclusion} then concludes the paper and discusses future work.

\section{Motivation}
\label{sec:motivation}

Two-way push-to-talk (PTT) radios, known as land-mobile radio (LMR), have been a staple of public safety communications for over eight decades. LMR systems represent the currently deployed mission-critical voice (MCV) communications solution. Unlike more modern cellular systems, LMR systems have been optimized to maximize the probability of successful transmission. The public safety community expects these systems to function consistently and reliably, particularly in challenging environments. 

The public safety community has considered transitioning the aging LMR infrastructure to broadband communications infrastructure. Before the transition can happen, it is necessary to understand the influence of key performance indicators (KPIs) such as network access time (PTT delay), mouth-to-ear delay (MtE), or codec performance under varying bit error rate (BER) on end-user quality of experience (QoE), and their relationship to task completion or success rate.


Establishing a relationship between KPIs and QoE is typically performed in human-subject studies in a controlled environment~\cite{voran2015speech}. During the study, the controlled environment (testbed) systematically varies the KPIs while the subject (study participant) uses the communication system emulated by the testbed to accomplish a communication-dependent task.  One a metric is chosen to measure QoE, the subject's accuracy, perceived QoE, and other data collected during the experiment can then be used to relate KPIs to QoE. Conducting such human-subject studies is time-consuming and logistically and administratively challenging. The software tools needed to create a testbed with sufficient accuracy and flexibility are generally unavailable.

Establishing appropriate initial KPI values is also challenging. If the resulting audio is too easy or hard for the study participants to understand, not much can be learned about the influence of the KPIs on QoE. Better methods are needed to establish initial KPI value ranges without relying on human feedback.


\section{Overall Study Purpose}
\label{sec:purpose}

The overall purpose of this study is to develop the methodology, tools, and public data sets to help the public safety community understand the influence of communication system parameters such as network access time (PTT delay), bit error rate (BER), background noise, and codecs on end-user quality of experience and their impact on the perceived and actual performance of tasks in a field environment.

We designed a testbed to emulate realistic channel and network conditions and used it to conduct a series of listening and interactive studies with human-subject volunteers. We then use the collected data to assess the influence of the selected impairment factors on the subjects' ability to understand the transmitted message.



The study investigates the influence of factors such as the voice codec, BER, and frame drop rate on the accuracy of a transcription task performed by the test subject.

\section{Experimental Setup and Procedures}
\label{sec:experimental-setup}


\subsection{Mission-Critical Voice Testbed}
\label{sec:experimental-setup:mcv-testbed}

\begin{figure}[ht]
  \centering
  \includegraphics[width=1\linewidth]{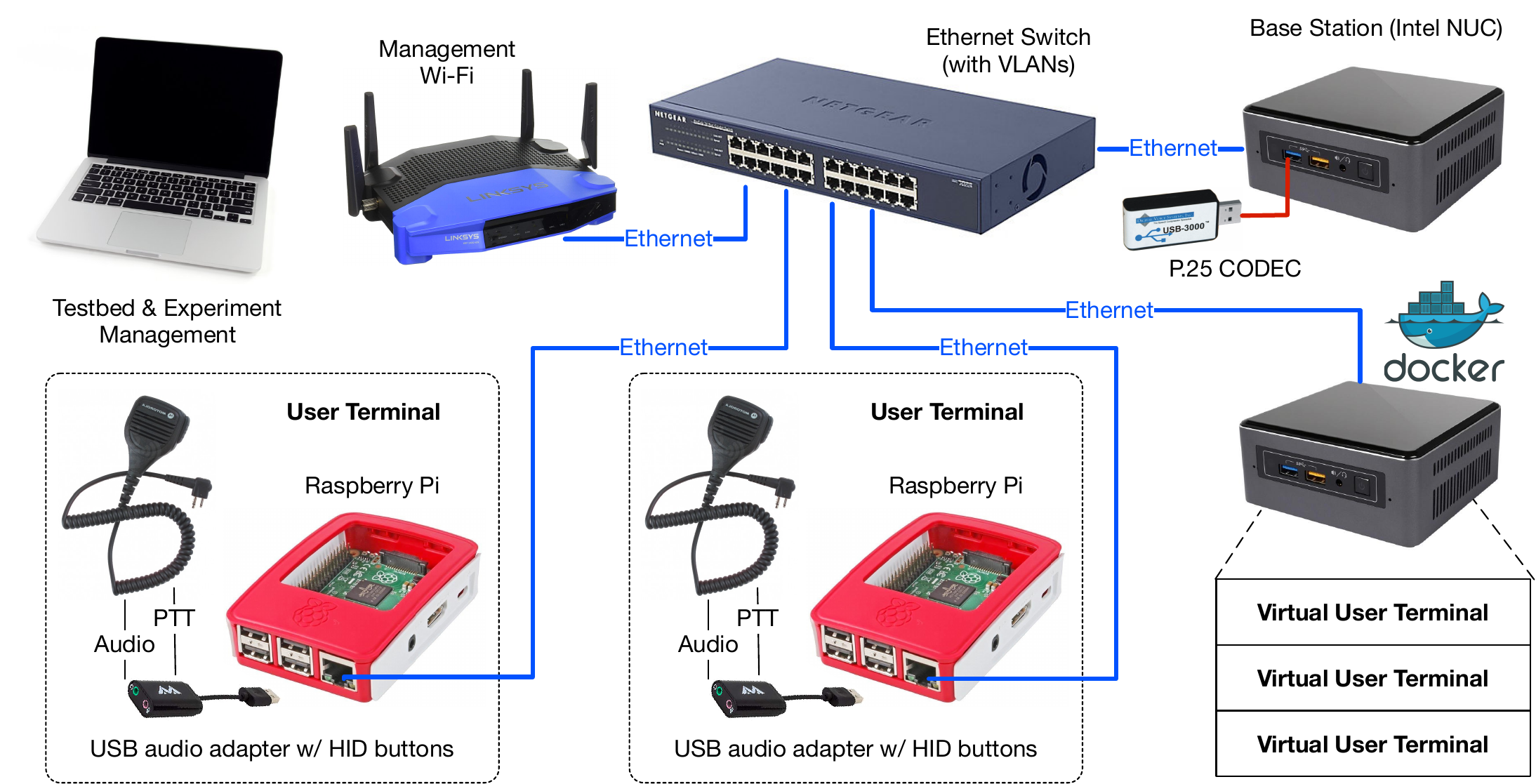}
  \caption{Mission-critical voice testbed architecture. The diagram shows a configuration with two user terminals.}
  \label{fig:testbed}
\end{figure}

The MCV testbed is a hardware and software architecture, shown in \cref{fig:testbed}, designed to support human-subject QoE studies with MCV. The testbed supports interactive two-way and listening one-way communication scenarios. In an interactive experiment, the participants communicate through the testbed's user terminals (shown in \cref{fig:testbed}) while trying to accomplish a task. In a listening experiment, the participants transcribe information from a series of impaired audio recordings as accurately as possible. The testbed emulates varying communication channel conditions by adjusting mouth-to-ear (MtE) delay, push-to-talk (PTT) button delay, codecs, bit error rate, background noise, and other parameters influencing end-user QoE.

\begin{figure}[ht]{%
    \setlength{\fboxsep}{0pt}%
    \setlength{\fboxrule}{1pt}%
    \subfloat{\fbox{\includegraphics[width=0.48\linewidth]{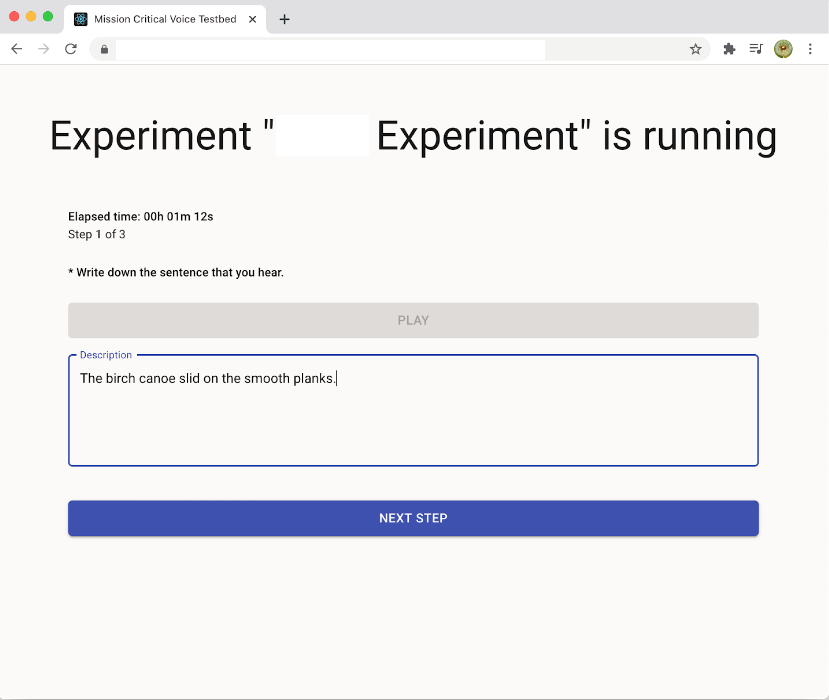}}}
    \hspace*{\fill}
    \subfloat{\fbox{\includegraphics[width=0.48\linewidth]{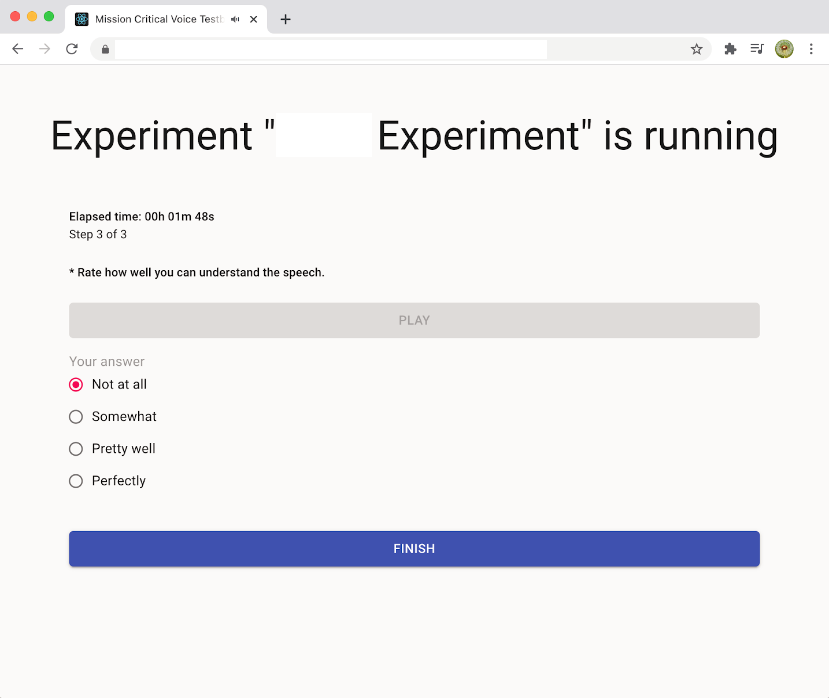}}}
    \caption{Testbed web user interface for listening experiments}\label{fig:listen-ui}
}\end{figure}

Listening experiment participants are sent a unique URL and access the experiment through a web-based user interface (UI), as shown in \cref{fig:listen-ui}. The UI presents a randomized series of impaired audio recordings. Each recording can only be played once. The participant is asked to extract some information from the recording as accurately as possible and enter it into the testbed's UI. The accuracy of the response will depend on the impairments applied to the audio recording. We then estimate the effects of the applied impairments on speech intelligibility by analyzing response accuracy.

The web-based UI runs in the participant's browser and communicates with the testbed through an HTTP-based API. The API provides access to the experiment metadata and audio recordings and is also used to submit responses entered by the participant.

\subsection{Automatic Speech Recognition Robot}
\label{sec:experimental-setup:asr-robot}

\begin{figure}[ht]{%
  \centering%
  \includegraphics[width=\linewidth]{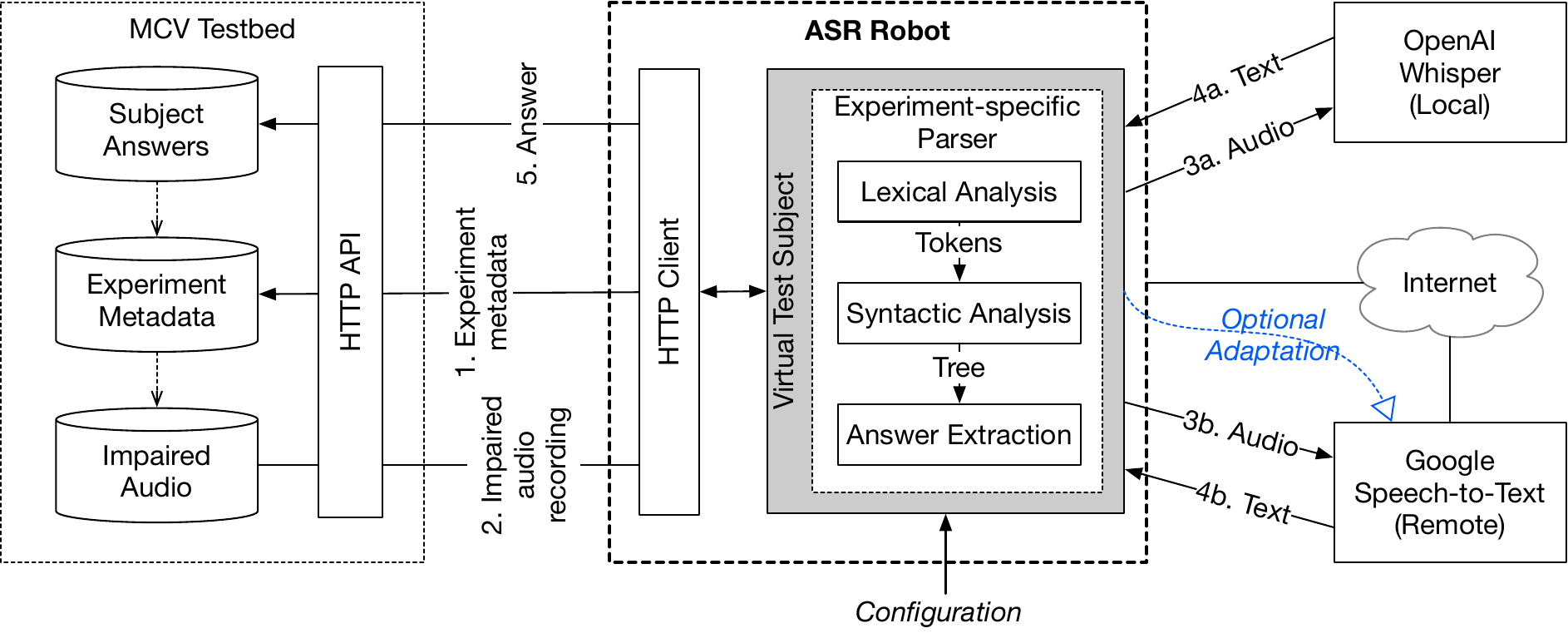}%
  \caption{Automatic speech recognition (ASR) robot architecture.}%
  \label{fig:asr-robot}
}\end{figure}

The automatic speech recognition (ASR) robot is a program that emulates human test subjects. The robot obtains experiment metadata and impaired audio recordings from the testbed, converts audio to text using either OpenAI Whisper~\cite{whisper} or Google Speech-to-Text (STT)~\cite{google-stt} engines, extracts the answer from the transcribed text, and submits the answer back to the testbed as a real human test subject would. \Cref{fig:asr-robot} illustrates the robot's architecture.

At a high level, the robot is a custom Python program coupled with a third-party ASR engine. The program communicates with the testbed via its HTTP REST API. This is the same API that powers the web UI, i.e., the robot has access to the same data as human test subjects and can also submit answers to the testbed through the same API. The robot can be configured to use either Whisper or Google STT. The robot runs the Whisper engine locally. When Google STT is selected, the robot interacts with the engine over the Internet. The transcribed text may contain errors in the form of missing or incorrect words.

The robot contains an internal experiment-specific text processing component (parser), which takes the transcribed text and produces an answer for the robot to submit back to the testbed. All our experiments involve New Jersey (NJ) license plate numbers encoded in the NATO phonetic alphabet~\cite{nato-alphabet}. Thus, we only implement a single parser for this purpose. The parser expects text with a known lead sentence followed by a NATO-encoded NJ license plate number. It extracts and returns the license plate number.

\subsubsection{Levenshtein Distance}

The Levenshtein Distance~\cite{levenshtein1966binary} is a metric for quantifying the difference between two strings. It is defined as the minimum number of insertions, deletions, and/or character substitutions required to change one string to the other. The Levenshtein distance is similar to other ``edit distance" type metrics such as the Hamming Distance; we found it is particularly well suited as a measure of QoE for the task at hand. With license experiments, we are less interested in testing for syntactical similarity and care more about phonetic similarity--the Levenshtein distance gives little penalty to strings that are similar in structure but ``shifted" by the addition or deletion of extra characters.

As detailed below, to compute our final QoE metric, we extract tokens of text with the ASR robot and measure the Levenshtein distance between these text tokens and NATO alphabet words.

\subsubsection{Measuring QoE}

The parser first splits the transcribed text into words using the Natural Language Toolkit (NLTK) library's~\cite{nltk} Treebank word tokenizer. Each transcription is supposed to start with the lead sentence ``Reporting license place''. The parser removes the lead sentence, any words with a Levenshtein distance smaller than three to the lead sentence words, and any words from the NLTK English stop word corpus. The remaining tokens represent a NJ license plate number encoded in the NATO phonetic alphabet.

Errors in the transcription process (due to the impairments) could produce tokens that cannot be directly mapped to NATO alphabet words. We use an approximate-matching algorithm that maps each token to the most likely NATO alphabet word. The algorithm first computes the BLEU score~\cite{bleu} (with the exponential decay smoothing method) and the Levenshtein distance between the token and each NATO phonetic alphabet word. The two metrics are converted to a score, and the algorithm chooses the NATO phonetic alphabet word with the best score. Equation~\ref{eq:argmax_p} formalizes this approach:

\begin{equation} \label{eq:argmax_p}
    SelectedWord = \arg \max_{x \in N}( score(x))
\end{equation}

where $N$ is the set of all NATO phonetic alphabet words and $score()$ is the chosen metric. A Levenshtein distance is converted to the score metric as follows:

\begin{equation}\label{eq:levenshtein_p}
\text{LevScore}(x) =  1- \frac {Levenshtein(x, token)} { \max(len(x), len(token))}
\end{equation}

where $x$ is a NATO alphabet word, and $token$ is one of the transcribed tokens (NATO-like words). The BLEU score is already in the $[0, 1]$ interval and can be directly plugged as the score into~\cref{eq:argmax_p}.

This approach resembles the human pattern-matching process, where a trained human test subject, in the case of uncertainty, selects the NATO word that is most similar to the word they hear. The robot then converts the list of NATO words back to letters and numbers and submits the result as an answer to the testbed. This process is repeated for every audio recording within the experiment.

The robot also calculates an overall experiment transcription accuracy score for each experimental run, which can be used to estimate the transcription confidence of the robot on the experiment. The overall score is calculated as the average accuracy over all impaired audio recordings within the experiment:

\begin{equation}
    \label{eq:experiment-score}
    \text{Experiment Score} =  \frac { S_{1} +S_{2} +S_{3} + ... + S_{n}} {n}
\end{equation}

where $S_{n}$ is a new per-recording score metric and $n$ is number of audio recordings within the experiment. We use two definitions for $S_n$, depending on whether the correct answer for each audio recording is known.

If the experiment metadata includes correct answers, $S_n$ is calculated as

\begin{equation}
    \label{eq:experiment-score-no-answers}
    S_{n} =  1- \frac {Levenshtein(number, correct)} { \max(len(number), len(correct))}
\end{equation}

where $number$ is the transcribed license plate number, and $correct$ is the correct answer (ground truth). If the experiment metadata does not include correct answers, $S_n$ is calculated as

\begin{equation}
S_n = \sum_{word \in token} LevScore(token)
\end{equation}

where $LevScore(token)$ is the Levenshtein-based score metric from \cref{eq:levenshtein_p}. Thus, when the correct answers are unknown, the overall experiment score is calculated as the average of the Levenshtein distance sums within each license plate, i.e., the similarity of the transcribed words to NATO words determines the overall experiment score.

\subsection{Brief Comparison of ASR Engines}
\label{sec:experimental-setup:asr-engines}

The ASR robot described in \cref{sec:experimental-setup:asr-robot} works with OpenAI Whisper~\cite{whisper} and Google Speech-to-Text~\cite{google-stt} engines. Both engines take audio recordings as input and produce a text-based transcription as output. This section briefly compares the two engines.

\subsubsection{Google Speech-to-Text (STT)}
\label{sec:experimental-setup:asr-engines:google-stt}

Google STT~\cite{google-stt} is a speech-to-text translation service provided by Google. The service is hosted in the Google cloud and requires a subscription. Access is provided through an API in the Google Cloud Platform~\cite{gcp}. The API supports streaming audio input and is capable of real-time translation. The service uses deep learning neural network algorithms with a proprietary model. This gives the application limited control over the used model version.

Until recently, Google STT was only available through GCP. Recently, Google has been offering an on-premises version of Google STT to trusted customers~\cite{stt-onprem}. This feature is intended for customers who need to process protected speech and must meet data residency and compliance requirements.

The Google STT service can be customized using domain-specific models or model adaptation. Domain-specific models are pre-trained for specific audio types, sources, or situations. The following models are available as of October 2023: long-form content, short utterances (commands or single-shot speech), video clips, phone calls, command and search, medical dictation, and conversation. Model adaptation allows the application to provide hints to help improve the accuracy of domain-specific terms and phrases. This feature also boosts certain words or phrases by assigning more weight to some phrases than others.

In all our experiments, we only use the cloud-based version of Google STT through the GCP API. We do not use domain-specific models or model adaptation.

\subsubsection{OpenAI Whisper}
\label{sec:experimental-setup:asr-engines:openai-whisper}

Whisper~\cite{whisper} is an open-source automatic speech recognition system developed by OpenAI. The system uses a Transformer-based deep learning neural network trained on more than 680,000 hours of multilingual and multitask supervised (paired with transcripts) data from the internet~\cite{whisper-intro}. Both the model and the inference program are open source.

Whisper comes in five model sizes (parameter number in parentheses): tiny (39\,M), base (74\,M), small (244\,M), medium (769\,M), and large (1550\,M). The size of the model (number of parameters) determines hardware requirements. Larger models generally provide better transcription results at the cost of running slower.

Whisper comes in the form of a Python library~\cite{whisper-github} that could be embedded into a custom application. The application then simply interacts with Whisper through the API exported by the package. We used this approach in the ASR robot described in~\cref{sec:experimental-setup:asr-robot}.

Whisper has been trained on 30-second segments and can only process audio in chunks of the same length. To transcribe longer audio segments, Whisper consecutively transcribes buffered 30-second chunks and implements a heuristic context window-shifting strategy.

The Python API does not support transcription streaming. An approximation of streaming could be built by repeatedly transcribing short chunks of audio, which comes with a performance penalty.


In all our experiments, we run Whisper with the medium and large model sizes. Running the large model with acceptable performance generally requires hardware acceleration (a GPU).

\subsection{Crowdsourcing Audio Transcriptions to Amazon Mechanical Turk}
\label{sec:experimental-setup:mturk}


Amazon Mechanical Turk (MTurk)~\cite{mturk} is a crowdsourcing marketplace that allows outsourcing human intelligence tasks to remote workers for a fee. Outsourcing multimedia tasks, e.g., audio transcription, to crowdsourcing services like MTurk has been proposed in the academic literature~\cite{hossfeld2014best,marge2010using,ntia2017:crowdsourced,ribeiro2011crowdmos}. The benefits of outsourcing tasks from a tightly controlled environment (lab) to MTurk workers include better repeatability, affordability, and generally shorter experiment completion times. The main drawbacks include the loss of control over the worker's (listening) environment and setup and significant worker (human subject) variability.

We use the MTurk service to recruit human test subjects for the listening experiment described in \cref{sec:experimental-setup:listening-design}. We designed a new listening experiment for MTurk workers in the testbed and generated a single link (HTTP URL) to be shared with all workers through the MTurk platform. Upon accepting the task, the worker opens the link, which takes them to the public-facing web UI of our testbed. The testbed first collects basic demographic information and asks the worker to sign a consent form. It then generates a unique set of impaired audio recordings for the worker. The testbed presents a simple UI where the worker individually listens to the audio recordings and enters the extracted information (NJ license plate number) into a web form.

\subsection{Listening Experiment Design}
\label{sec:experimental-setup:listening-design}

The listening experiment consists of impaired audio recordings of NJ license plate numbers encoded with the NATO phonetic alphabet~\cite{nato-alphabet}. The goal is to record the license plate number in the original (non-NATO) form as accurately as possible.

The experiment consist of 60 audio recordings generated with the Google Text-to-Speech (TTS) engine. A female voice is used for all recordings. Each audio recording begins with the lead sentence `Reporting license plate'' followed by an NJ license plate number encoded with the NATO phonetic alphabet. A NJ license plate number consists of one letter, followed by two digits, followed by three letters. We generate the license plate numbers randomly using a custom tool~\cite{mcv-testbed}.

Each audio recording is encoded using the P.25 Phase 2 or AMR wideband codec. The codec bit stream is then impaired with bit-error bursts according to the Gilbert-Elliot model~\cite{ge1,ge2} for correlated bit error patterns. We keep the state transition probabilities constant ($P_{GB}=0.01$, $P_{BG}=1-P_{GB}$) and only vary the burst size $k$ using the following values: 1, 2, 4, 6, 8, 10. Each audio recording is randomly assigned a combination of the two experimental conditions. The corrupted bit stream is then decoded back to raw audio data in the WAVE format.

All test subjects receive the same non-repeating license plates in a random order. The conditions vary across the recordings. For example, one participant may receive a recording encoded with the AMR wideband codec with $Pgb=0.01$, $K=2$, whereas another participant may receive the same recording encoded with the P.25 Phase 2 codec with $Pgb=0.01$, $K=10$.

Twenty-nine participants recruited via the Amazon Mechanical Turk service were human test subjects. The participants completed 1,740 trials (60 each); three trials were excluded due to an audio playback issue, resulting in 1,737 trials across all participants.

Following the human-subject trials, the ASR robot was also used to transcribe the same set of impaired audio recordings for comparison. We ran the robot three times with Google STT, Whisper Medium, and Whisper Large engines. The analytic sample consists of 3,584 trials across the four participant groups: humans, Google Speech-to-Text, Whisper Medium, and Whisper Large.

\section{Analytical Procedures}
\label{sec:analytical-procedures}


We compare the trial transcriptions to the correct answers to measure performance and then calculate two accuracy indicators. First, we use correctness as a measure of overall accuracy. We graph the experiment scores for each experiment run to visually inspect and descriptively compare the performance of each subject type (i.e., human test subjects and the three ASR system variations).

Second, we use a continuous indicator to measure the degree of accuracy, which is important to consider, given that even partial license plates can benefit emergency response situations. As such, we use the Levenshtein distance \cite{levenshtein1966binary}, which is a measure of the difference between two strings and reflects the minimum number of single-character edits necessary to change one string to another string (e.g., correct response). In this application, the distance reflects the number of single-character transformations needed to change an incorrect response to the correct response. Because the maximum number of digits/characters on NJ license plates is six, we truncate the Levenshtein distances at a maximum of six. 

We assess these measures across experimental conditions and between subject types. More specifically, we compare humans' and ASR systems' responses for each combination of codec and bit error burst size experimental conditions to assess whether (1) ASR robots perform significantly worse or better than humans and (2) under which experimental conditions are the differences between humans and ASR systems largest.

For model parsimony, we analyze the P.25 Phase 2 and AMR wideband trials separately for each pairwise comparison of interest (i.e., humans vs. Google Speech-to-Text,  humans vs. Whisper Medium, and humans vs. Whisper Large).  We regress Levenshtein distance onto bit error manipulation and subject type in linear regression models. We add a multiplicative interaction term (bit error burst size x subject type) to assess whether accuracy differs between humans and ASR systems at each bust size (within each codec). We decompose the interaction to examine pairwise comparisons between the interest groups to determine whether differences are statistically significant. Because trials are nested within human participants or ASR systems, we account for correlated standard errors by estimating these models using cluster robust standard errors. Additionally, given the large number of pairwise comparisons examined (38), we apply a Benjamini-Hochberg correction \cite{BHadjustment} to minimize the likelihood of Type I errors by controlling the expected proportion of falsely rejected hypotheses (i.e., the false discovery rate).
\section{Results}
\label{sec:results}





\subsection{Levenshtein Distance}

\begin{figure*}[ht]
  \centering
  \includegraphics[width=1\linewidth]{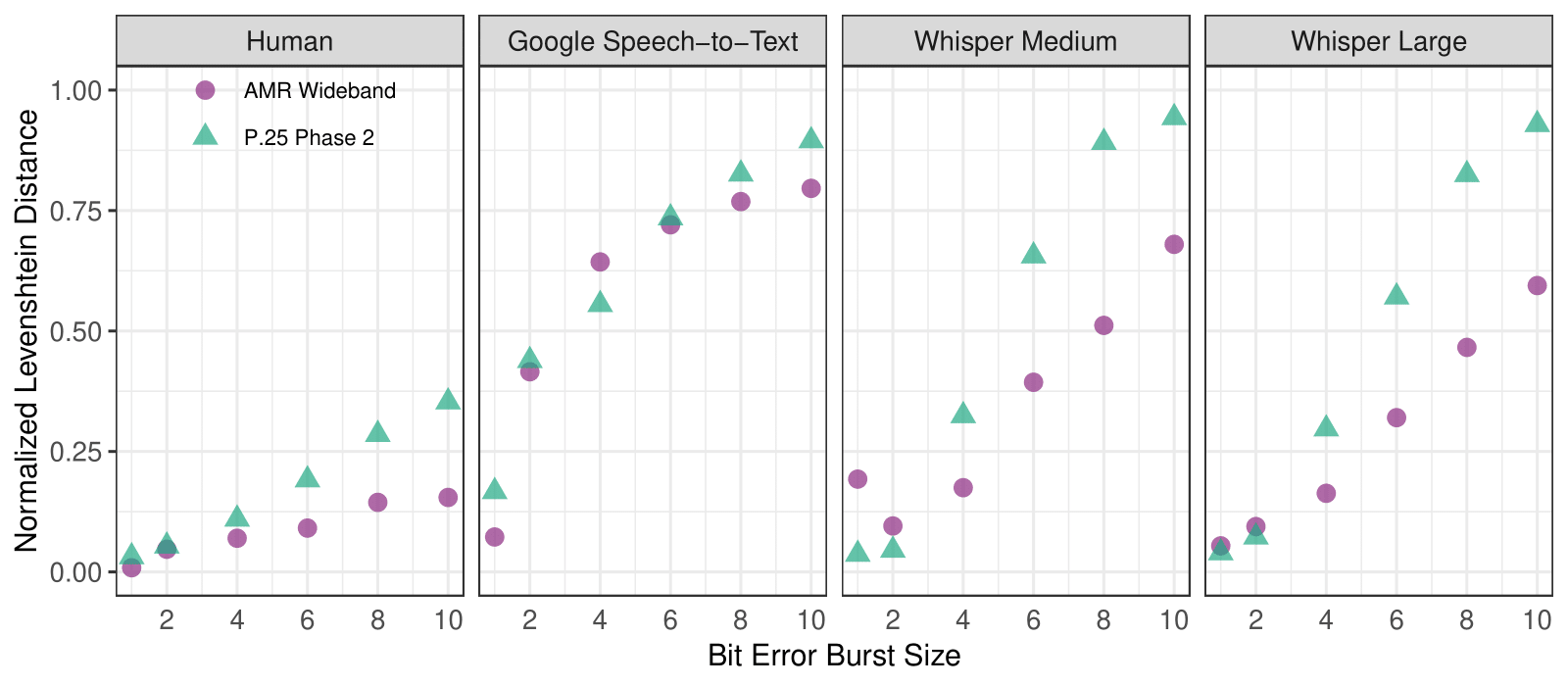}
  \caption{Normalized Levenshtein distance of answers as a function of experimental conditions (codec and bit error burst size). K=1 was under Pgb=0.00, P=0.00 and all other bit error burst sizes were under Pgb=0.01, P=0.01.}
  \label{fig:results:levenshtein-distance}
\end{figure*}

Normalized Levenshtein distances are compared between the subject types to assess how the ASR systems perform across the varied experimental conditions of codec and bit error burst size. \Cref{fig:results:levenshtein-distance} shows the average Levenshtein distance of answers (normalized for graphing purposes) broken down by codec and bit error burst size, with lower scores indicating greater accuracy. Pairwise comparisons show significant differences ($p < .001$) between humans and Google Speech-to-Text under all experimental conditions, suggesting that humans perform significantly better than Google Speech-to-Text regardless of codec and bit-error bust-size variations.

Comparatively, there is more significant variation in performance between the ASR systems and humans. Under the AMR wideband codec and $K = 1$ burst size, humans perform better than Whisper Large, but not to a statistically significant degree (95\% CI [-0.12, 0.04], $p = .345$). Under the P.25 Phase 2 codec and $K = 1$ burst size, both Whisper Medium  (95\% CI [-0.04, 0.16], $p = .274$) and Whisper Large (95\% CI [-0.04, 0.16], $p = .277$) perform better than humans, but not to statistically significant degrees. In all other pairwise comparisons, humans perform significantly better than the ASR systems ($p < .001$). Interestingly, the Whisper ASR systems performed better under the P.25 Phase 2 codec for the two lowest bit error burst sizes ($K = 1$ and $K = 2$) but better under the AMR wideband codec for burst sizes $K = 4$ and larger. In contrast, humans and Google Speech-to-Text perform better under the AMR wideband codec for all burst sizes.

Given the focus on understanding quality of experience, we also assessed how humans' accuracy varied within codecs. Under the P.25 Phase 2 codec, human performance consistently declined across the consecutive bit error burst sizes to a statistically significant degree ($p < .050$).  In contrast, under AMR wideband,  although humans performed significantly better at $K = 1$ than $K = 2$ (95\% CI [-0.39, -0.07], $p = .005$) and $K = 6$ than $K = 8$ (95\% CI [-0.52, -0.12], $p = .002$), the difference in performance was not significantly different between $K = 2$ and $K = 4$ (95\% CI [-0.31,  0.04], $p = .124$), $K = 4$ and $K = 6$ (95\% CI [-0.39, -0.07], $p = .128$), and $K = 8$ and $K = 10$ (95\% CI [-0.31,  0.19], $p = .626$). As such, AMR wideband appears to be more robust to bit error, with only two areas of significant reduction in quality of experience identified by this analysis.

\subsection{Experiment Scores}
\begin{figure}
    \centering
    \includegraphics[width=1\linewidth]{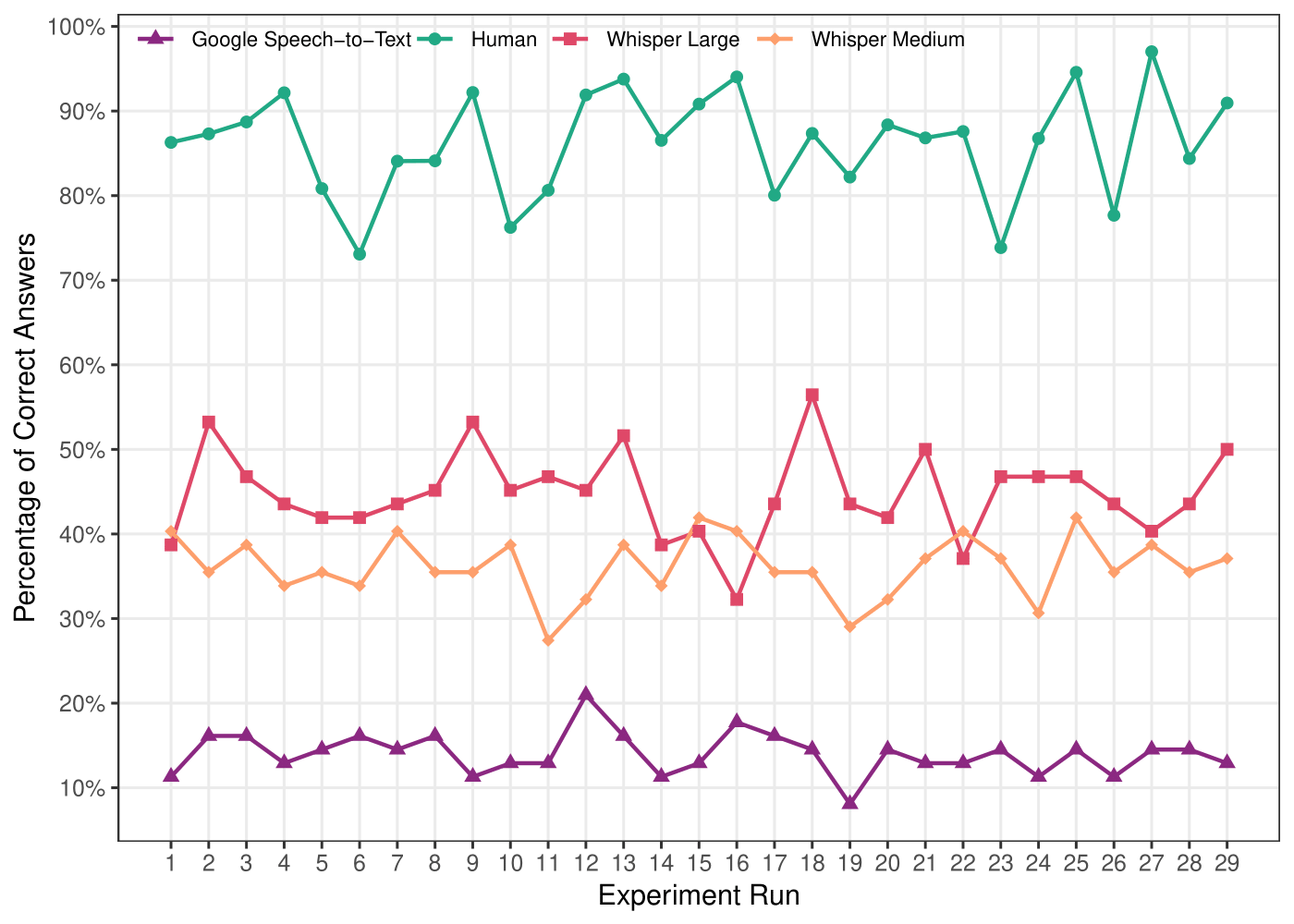}
    \caption{A comparison of overall experiment scores, converted to percentages, across twenty-nine experiment runs, one for each human test subject. The vertical axis allows a comparison of the relative performance of humans against three ASR robot variations based on the Google Speech-to-Text engine and two versions of the OpenAI Whisper engine.}
    \label{fig:results:experiment-score}
\end{figure}

Experiment scores are compared between the subject types to assess how the ASR systems perform on each experiment run compared to the respective human, with higher scores indicating greater accuracy.  As shown in \cref{fig:results:experiment-score}, humans perform best across all runs, whereas Google Speech-to-Text performs the worst. Whisper Large generally performs better than Whisper Medium, with Whisper Large having higher experiment scores than Whisper Medium in 25 of the 29 runs.

\section{Interactive Experiment Design}
\label{sec:experimental-setup:interactive-design}

\begin{figure}[ht]
  \centering
  \includegraphics[width=\linewidth]{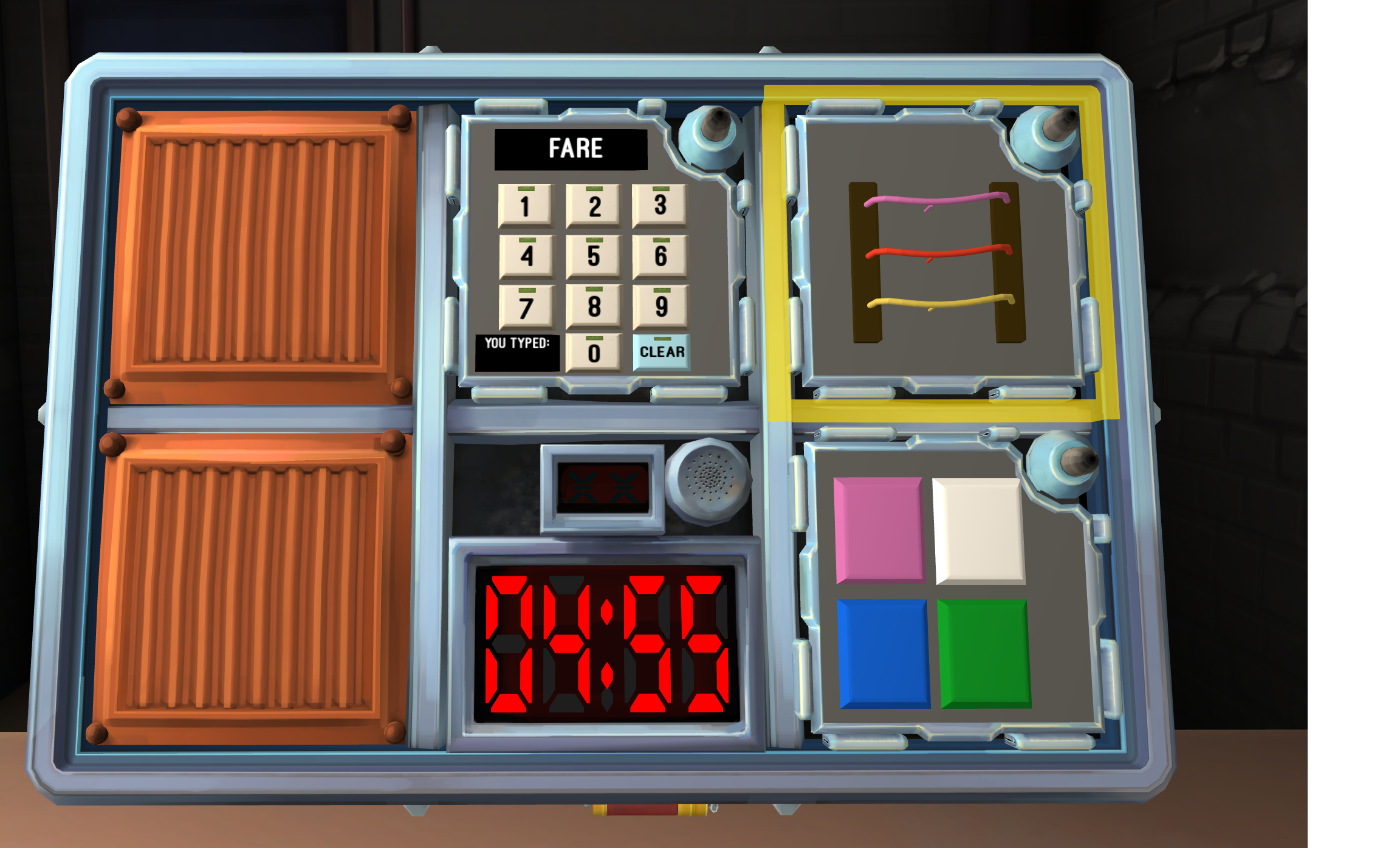}
  \caption{Example of "bomb" interface for interactive experiments with {\em Keep Talking and Nobody Explodes}.}
  \label{fig:keeptalking}
\end{figure}

In line with the goal of further improving the emulation of challenging, real-world critical missions, we also made an effort to design interactive human-subject experiments centered around the commercially available computer game {\em Keep Talking and Nobody Explodes}, developed by Steel Crate Games in 2015. We modified the game using the {\em Unity} development kit. In this section, we briefly summarize the game and our experimental design.

\subsubsection{Game Summary and Objective}
\label{sec:experimental-setup:interactive-design:objective}

Keep Talking and Nobody Explodes is a cooperative game for two communicating human players. One player manipulates a 3D depiction of a ``bomb'' (\cref{fig:keeptalking}) on a terminal. The second player has a "bomb manual" but cannot view the terminal (bomb). The player at the terminal describes the visual aspects of the bomb to the player with the manual; the player with the manual communicates instructions to defuse the bomb to the first player. Cooperation and the ability to accurately communicate and comprehend instructions are crucial to accomplishing the task.

\subsubsection{Experiment Design}
\label{sec:experimental-setup:interactive-design:design}

We used the MCV testbed (\cref{sec:experimental-setup:mcv-testbed}) for all communication between the players. We recruited participants from Columbia University and the New York City metropolitan area to perform the {\em defuser} role. A trained FDNY dispatcher performed the manual role ({\em dispatcher}). The defuser and dispatcher were in separate rooms and could only communicate through the testbed. The experimenter independently varied the bit error rate (BER) of the communication channel at the start of each trial.

We created five distinct bomb configurations for the experiment, one for each trial. Each test subject receives the same configurations. Each bomb configuration contained three module types (shown in \cref{fig:keeptalking}):
\begin{itemize}
    \item "wires"---the module has 2--5 wires of different colors; the defuser must select the wires in the correct sequence.
    \item "buttons"---the module 2--5 buttons of different colors and configurations; the defuser must press buttons in the correct sequence.
    \item "keypad"---a numerical keypad with a word displayed over the buttons is shown; the defuser must type the correct keypad sequence.
\end{itemize}

The defuser has 5 minutes to defuse each bomb configuration. In each trial, the defuser signals to the dispatcher when they are ready to start, and the dispatcher acknowledges, at which point the timer begins. The defuser and dispatcher communicate back and forth through the terminals to defuse the bomb. There is only one correct way to defuse each bomb module. The defuser must carefully follow the dispatcher's instructions. An incorrect interaction with the bomb, e.g., pressing the wrong button, counts as a strike. The bomb explodes after three strikes or when the timer runs out. A trial is successful when the defuser disables all three modules before the time limit.

We record the communication impairments configured in the testbed and the total time to complete the trial in each experiment.

\section{Discussion}
\label{sec:discussion}



Our system can utilize both Google Speech-to-Text and OpenAI Whisper speech recognition engines to perform transcription of impaired audio recordings of NJ license plate numbers, and we compare their performance with the performance of human subjects hired through Amazon Mechanical Turk.

Our findings reveal that, in general, the human test subjects perform better than all tested ASR robot configurations. In terms of both percentage accuracy and the "Experiment Score" metric, the medium and large Whisper models can outperform Google Speech-to-Text. We also observe differences in performance across codecs. Humans perform best on AMR wideband (vs. P.25 Phase 2) tasks across all burst sizes, but this accuracy gap is most pronounced at larger burst sizes. This intelligibility difference suggests that the codec significantly contributes to humans' QoE when performing mission-critical tasks.

We acknowledge some key limitations in this study. Assessing the quality of the ASR system relative to human test subjects requires running experiments with human test subjects themselves, which brings several practical and logistical concerns. Scheduling time for subjects to participate, as well as the overhead in actual experimental setup and execution, could contribute to being a significant time barrier. The data sets we used for ASR testing are also relatively small, which could have influenced the overall performance of those engines.

Additionally, accuracy is the only measure examined in this study. While accuracy is an important indicator of intelligibility and audio quality, other KPIs are also important when considering QoE, such as latency and end-to-end access time. As such, future replications of this work should incorporate other measures for a more robust understanding of QoE.\cite{NIST:QoE}

We originally planned to conduct the experiments through in-person two-way communication sessions with trained first responders in a controlled environment on the Columbia University campus. The onset of the COVID-19 pandemic forced us to reconsider. We perform a series of listening experiments instead, first remotely with the human subjects accessing the testbed from home, then in-person in a lab on the campus. Due to the difficulties of recruiting trained first responders, most of our test subjects were Columbia University students or general-population volunteers.

The pandemic and the general difficulty of finding suitable test subjects also prompted us to consider using the ASR robot and Amazon MTurk~\cite{mturk}. We use MTurk to recruit a larger human test subject population for our listening experiments. To our knowledge, this is the first study using ASR and MTurk to investigate MCV communications systems.



\section{Related Work}
\label{sec:related-work}

Outsourcing multimedia tasks to platforms such as Amazon Mechanical Turk (MTurk)~\cite{mturk} is popular~\cite{hossfeld2014best}. Marge et al. investigate the use MTurk as a method for transcription of spoken language~\cite{marge2010using}. The study found that MTurk transcriptions are generally accurate. The authors also show how accuracy could be improved by combining transcriptions from several MTurk workers using word-level voting according to the NIST ROVER algorithm~\cite{rover}. Voran and Catellier~\cite{ntia2017:crowdsourced} propose the Crowdsourced Modified Rhyme Test (CMRT), a speech intelligibility test based on the Amazon MTurk platform which produces results comparable to the laboratory version of MRT. The authors find CMRT more repeatable and affordable. CrowdMOS~\cite{ribeiro2011crowdmos} is a collection of tools to perform mean opinion score (MOS) tests on MTurk.

NIST PSCR has developed methods to measure and quantify various MCV key quality of experience (QoE) parameters, including mouth-to-ear (MtE) latency~\cite{pscr2018:mte-methods} and the time needed to establish a talk path~\cite{pscr2019:access-time-methods,pscr2020:acess-time-addendum,pscr2021:start-of-word}. Both methods assume a physical system under test (SUT), i.e., existing MCV communication terminals. The two methods inspire the design of our testbed. In contrast, we emulate the target MCV communication system in software with configurable system parameters rather than relying on physical MCV terminals.

Atkinson and Catellier conducted a human-subject study to evaluate the intelligibility of two analog FM and two digital P.25 systems in high-background-noise environments experienced by firefighters~\cite{atkinson2008intelligibility}. The study found analog FM systems to have higher intelligibility than digital systems. In some environments, none of the systems perform well. A follow-up study~\cite{ntia2013:p25-noise} found that updated P.25 systems perform better in noisy environments than analog FM systems. Several PSCR studies~\cite{ntia2012:amr-intelligibility,voran2015codecs,ntia2015:codecs-lte,ntia2017:codecs} evaluate speech intelligibility of various digital codecs in different acoustic noise and channel error environments using the Modified Rhyme Test (MRT). It is also desirable for MCV systems to transmit secondary information such as speaker identification or stress. Several PSCR studies~\cite{ntia2009:speaker-identification,voran2008listener,catellier2008speaker} characterize the relationship between speech intelligibility, speaker identification, and speaker stress in simulated speech processing conditions. The studies mentioned in this paragraph inspire the design of the listening experiments described in~\cref{sec:experimental-setup:listening-design}.

Mauch and Ewert propose the Audio Degradation Toolbox (ADT) for controlled degradation of audio signals~\cite{mauch2013audio}. The ADT provides configurable degradation units (e.g., add noise, add sound, high-pass filter, etc.) and a framework for composing (chaining) the degradations. The toolbox also provides profiles for common degradations such as smartphone playback, radio broadcast, or vinyl. Like in the ADT, the design of the audio degradation subsystem in our MCV testbed is composable. The testbed provides a broader selection of audio degradation components, focuses on MCV applications, and can degrade the audio signal in real time.



\section{Conclusion and Future Work}
\label{sec:conclusion}

We designed and implemented software tools for human-subject experiments with MCV. The tools include a testbed for designing, conducting, and evaluating human-subject experiments with MCV (\cref{sec:experimental-setup:mcv-testbed}) and an ASR robot (\cref{sec:experimental-setup:asr-robot}) designed to emulate humans in listening experiments. We conducted a transcription experiment where human subjects extract NJ license plate numbers from transcoded and impaired audio recordings. We evaluated the influence of correlated bit errors on the intelligibility of audio transcoded with P.25 Phase 2 and AMR wideband codecs. We also assessed the performance of the ASR robot on such audio.

We used an accuracy indicator based on the Levenshtein distance to compare the performance of humans and the ASR robot. We compared humans' and the robot's responses for each combination of codec and bit error burst size to asses (1) whether the ASR robot performs significantly worse or better than humans; and (2) under which experimental conditions are differences the largest. Humans performed the best across all experiment runs. The ASR robot based on Google Speech-to-Text (STT) performed the worst. The Whisper Large-based ASR robot performed better than the Whisper Medium model in 25 of 29 runs. A pairwise comparison of Levenshtein distances showed that humans performed significantly better than Google STT regardless of codec and bit error size variations. In audio transcoded with the P.25 Phase 2 codec, the Whisper-based ASR robot performed better than humans on short bit-error bursts. The difference between human and ASR performance grows larger with increasing burst size. These findings suggest that the codec significantly contributes to humans' QoE when performing MCV tasks.



We plan to better characterize the ``true'' performance of the ASR robot compared to humans in the future. We also intend to generalize our findings to different languages and accents within the same language. These investigations will contribute to a more comprehensive understanding of the applicability of ASR for MCV-related tasks. We will expand the scope of our experiments to include additional settings and real-life situations. We aim to evaluate the performance of the ASR system in challenging environments, such as high-noise urban areas, adverse weather conditions, and recordings with low audio quality.

We will also consider enhancing the overall accuracy of the ASR system. This entails fine-tuning the model using targeted training data specific to license plate recognition, optimizing the architecture to handle specific linguistic and phonetic characteristics of license plate numbers, and implementing advanced noise reduction techniques to improve performance in noisy environments.

\section*{Software Artifacts}

The source code for the MCV testbed and the ASR robot is available\footnote{\url{https://gitlab.com/irtlab/mcv-testbed}} under a permissive open-source license. Experimental impaired audio recordings and anonymized human-subject answer data are also available~\footnote{\url{https://gitlab.com/irtlab/mcv-experiments}}.

\section*{Acknowledgment}
\addcontentsline{toc}{section}{Acknowledgment}
We thank Jose Daniel Rubenstein for creating the initial prototype of the ASR robot.

We are grateful to Artiom Baloian for developing the listening experiment generator tool, organizing the listening experiments, and recruiting and supervising the human subject volunteers involved in the study.

We thank Luoyao Hao, Kunal Mahajan, Ian Loeb, Shaun Maher, and Michael Paciullo for their help with the testbed implementation.

Finally, we thank the NIST PSCR team for helpful suggestions, fruitful discussions, advice, and guidance.

This work was funded by NIST under the Federal Award ID 70NANB19H003.

\bibliographystyle{IEEEtran}
\bibliography{bibs/references,bibs/janakj,bibs/ietf,bibs/software,bibs/intelligibility,bibs/pscr,bibs/ntia}

\begin{thebibliography}{10}
\providecommand{\url}[1]{#1}
\csname url@samestyle\endcsname
\providecommand{\newblock}{\relax}
\providecommand{\bibinfo}[2]{#2}
\providecommand{\BIBentrySTDinterwordspacing}{\spaceskip=0pt\relax}
\providecommand{\BIBentryALTinterwordstretchfactor}{4}
\providecommand{\BIBentryALTinterwordspacing}{\spaceskip=\fontdimen2\font plus
\BIBentryALTinterwordstretchfactor\fontdimen3\font minus
  \fontdimen4\font\relax}
\providecommand{\BIBforeignlanguage}[2]{{%
\expandafter\ifx\csname l@#1\endcsname\relax
\typeout{** WARNING: IEEEtran.bst: No hyphenation pattern has been}%
\typeout{** loaded for the language `#1'. Using the pattern for}%
\typeout{** the default language instead.}%
\else
\language=\csname l@#1\endcsname
\fi
#2}}
\providecommand{\BIBdecl}{\relax}
\BIBdecl

\bibitem{funding}
\BIBentryALTinterwordspacing
{NIST Public Safety Innovation Accelerator Program}. (2018) {Mission Critical
  Voice Quality of Experience -- Notice of Funding Opportunity}. [Online].
  Available:
  \url{https://www.nist.gov/system/files/documents/2018/05/29/2018-nist-psiap-mcvqoe_nofo.pdf}
\BIBentrySTDinterwordspacing

\bibitem{dumas1999practical}
J.~S. Dumas and J.~C. Redish, \emph{A practical guide to usability
  testing}.\hskip 1em plus 0.5em minus 0.4em\relax Intellect Books, 1999.

\bibitem{levenshtein1966binary}
V.~I. Levenshtein, ``Binary codes capable of correcting deletions, insertions,
  and reversals,'' in \emph{Soviet Physics Doklady}, vol.~10, no.~8, 1966, pp.
  707--710.

\bibitem{voran2015speech}
S.~Voran and A.~Catellier, ``Speech codec intelligibility testing in support of
  mission-critical voice applications for lte,'' \emph{NTIA Report}, pp.
  15--520, 2015.

\bibitem{whisper}
A.~Radford, J.~W. Kim, T.~Xu, G.~Brockman, C.~McLeavey, and I.~Sutskever,
  ``Robust speech recognition via large-scale weak supervision,'' \emph{arXiv
  preprint arXiv:2212.04356}, 2022.

\bibitem{google-stt}
\BIBentryALTinterwordspacing
Google. Google speech-to-text. [Online]. Available:
  \url{https://cloud.google.com/speech-to-text}
\BIBentrySTDinterwordspacing

\bibitem{nato-alphabet}
\BIBentryALTinterwordspacing
{NATO} phonetic alphabet. [Online]. Available:
  \url{https://en.wikipedia.org/wiki/NATO_phonetic_alphabet}
\BIBentrySTDinterwordspacing

\bibitem{nltk}
\BIBentryALTinterwordspacing
{NLTK Project}. {Natural Language Toolkit (NLTK)}. [Online]. Available:
  \url{https://www.nltk.org/}
\BIBentrySTDinterwordspacing

\bibitem{bleu}
\BIBentryALTinterwordspacing
K.~Papineni, S.~Roukos, T.~Ward, and W.-J. Zhu, ``{BLEU}: A method for
  automatic evaluation of machine translation,'' in \emph{Proceedings of the
  40th Annual Meeting on Association for Computational Linguistics}, ser. ACL
  '02.\hskip 1em plus 0.5em minus 0.4em\relax USA: Association for
  Computational Linguistics, 2002, p. 311–318. [Online]. Available:
  \url{https://doi.org/10.3115/1073083.1073135}
\BIBentrySTDinterwordspacing

\bibitem{gcp}
\BIBentryALTinterwordspacing
{Google}. {Google Cloud Platform}. [Online]. Available:
  \url{https://cloud.google.com}
\BIBentrySTDinterwordspacing

\bibitem{stt-onprem}
\BIBentryALTinterwordspacing
------. {Google Speech-to-Text On-Prem}. [Online]. Available:
  \url{https://cloud.google.com/speech-to-text/priv}
\BIBentrySTDinterwordspacing

\bibitem{whisper-intro}
\BIBentryALTinterwordspacing
{OpenAI}. {Introducing Whisper}. [Online]. Available:
  \url{https://openai.com/research/whisper}
\BIBentrySTDinterwordspacing

\bibitem{whisper-github}
\BIBentryALTinterwordspacing
------. {Whisper GitHub Repository}. [Online]. Available:
  \url{https://github.com/openai/whisper}
\BIBentrySTDinterwordspacing

\bibitem{mturk}
\BIBentryALTinterwordspacing
{Amazon}. (2023) {Amazon Mechanical Turk}. [Online]. Available:
  \url{https://www.mturk.com}
\BIBentrySTDinterwordspacing

\bibitem{hossfeld2014best}
\BIBentryALTinterwordspacing
T.~Hossfeld, M.~Hirth, J.~Redi, F.~Mazza, P.~Korshunov, B.~Naderi, M.~Seufert,
  B.~Gardlo, S.~Egger, and C.~Keimel, ``{Best Practices and Recommendations for
  Crowdsourced QoE - Lessons learned from the Qualinet Task Force
  ''Crowdsourcing''},'' Oct. 2014, lessons learned from the Qualinet Task Force
  ''Crowdsourcing'' COST Action IC1003 European Network on Quality of
  Experience in Multimedia Systems and Services (QUALINET). [Online].
  Available: \url{https://hal.science/hal-01078761}
\BIBentrySTDinterwordspacing

\bibitem{marge2010using}
M.~Marge, S.~Banerjee, and A.~I. Rudnicky, ``Using the {Amazon Mechanical Turk}
  for transcription of spoken language,'' in \emph{2010 IEEE International
  Conference on Acoustics, Speech and Signal Processing}.\hskip 1em plus 0.5em
  minus 0.4em\relax IEEE, 2010, pp. 5270--5273.

\bibitem{ntia2017:crowdsourced}
\BIBentryALTinterwordspacing
S.~D. Voran and A.~A. Catellier, ``A crowdsourced speech intelligibility test
  that agrees with, has higher repeatability than, lab tests,'' National
  Telecommunications and Information Administration, Tech. Rep. TR-17-523, Feb.
  2017. [Online]. Available:
  \url{https://its.ntia.gov/publications/download/TM-17-523.pdf}
\BIBentrySTDinterwordspacing

\bibitem{ribeiro2011crowdmos}
F.~Ribeiro, D.~Flor{\^e}ncio, C.~Zhang, and M.~Seltzer, ``{CrowdMOS}: An
  approach for crowdsourcing mean opinion score studies,'' in \emph{2011 IEEE
  international conference on acoustics, speech and signal processing
  (ICASSP)}.\hskip 1em plus 0.5em minus 0.4em\relax IEEE, 2011, pp. 2416--2419.

\bibitem{mcv-testbed}
\BIBentryALTinterwordspacing
J.~Janak, A.~Baloian, D.~Rubenstein, and H.~Schulzrinne. A platform for
  experimental research in mission-critical voice. [Online]. Available:
  \url{https://gitlab.com/irtlab/mcv-testbed}
\BIBentrySTDinterwordspacing

\bibitem{ge1}
E.~N. Gilbert, ``Capacity of a burst-noise channel,'' \emph{The Bell System
  Technical Journal}, vol.~39, no.~5, pp. 1253--1265, 1960.

\bibitem{ge2}
E.~O. Elliott, ``Estimates of error rates for codes on burst-noise channels,''
  \emph{The Bell System Technical Journal}, vol.~42, no.~5, pp. 1977--1997,
  1963.

\bibitem{BHadjustment}
Y.~Benjamini and Y.~Hochberg, ``Controlling the false discovery rate: A
  practical and powerful approach to multiple testing,'' \emph{Journal of the
  Royal Statistical Society: Series B (Methodological)}, vol.~57, no.~1, pp.
  289--300, 1995.

\bibitem{NIST:QoE}
J.~Pieper, J.~Frey, and G.~Howarth, ``Mission critical voice quality of
  experience probability of successful delivery measurement methods,'' National
  Institute of Standards and Technology, Tech. Rep., 2023.

\bibitem{rover}
J.~G. Fiscus, ``{A Post-Processing System to Yield Reduced Word Error Rates:
  Recognizer Output Voting Error Reduction (ROVER)},'' in \emph{1997 IEEE
  Workshop on Automatic Speech Recognition and Understanding
  Proceedings}.\hskip 1em plus 0.5em minus 0.4em\relax IEEE, 1997, pp.
  347--354.

\bibitem{pscr2018:mte-methods}
J.~Frey, J.~Pieper, and S.~Thompson, ``Mission critical voice {QoE}
  mouth-to-ear latency measurement methods,'' National Institute of Standards
  and Technology, {Gaithersburg, MD}, Tech. Rep. NIST Interagency/Internal
  Report (NISTIR) 8206, Mar. 2018.

\bibitem{pscr2019:access-time-methods}
J.~Pieper, J.~Frey, C.~Greene, Z.~Soetan, S.~Thompson, D.~Bradshaw, and
  S.~Voran, ``Mission critical voice quality of experience access time
  measurement methods,'' National Institute of Standards and Technology,
  Gaithersburg, MD, Tech. Rep. {NIST Interagency/Internal Report (NISTIR)
  8275}, Oct. 2019.

\bibitem{pscr2020:acess-time-addendum}
C.~Greene, J.~Frey, Z.~Soetan, J.~Pieper, and S.~Thompson, ``Mission critical
  voice quality of experience access time measurement method addendum,''
  Gaithersburg, MD, Dec. 2020.

\bibitem{pscr2021:start-of-word}
\BIBentryALTinterwordspacing
W.~Magrogan, J.~Pieper, and Z.~Soetan, ``Mission critical voice start-of-word
  correction for access delay measurement system,'' National Institute of
  Standards and Technology, Gaithersburg, MD, Tech. Rep. Technical Note (NIST
  TN) 2166, Jun. 2021. [Online]. Available:
  \url{https://tsapps.nist.gov/publication/get_pdf.cfm?pub_id=932595}
\BIBentrySTDinterwordspacing

\bibitem{atkinson2008intelligibility}
D.~J. Atkinson and A.~A. Catellier, \emph{Intelligibility of selected radio
  systems in the presence of fireground noise: Test plan and results}.\hskip
  1em plus 0.5em minus 0.4em\relax US Department of Commerce, National
  Telecommunications and Information Administration, 2008.

\bibitem{ntia2013:p25-noise}
\BIBentryALTinterwordspacing
------, ``Intelligibility of analog {FM} and updated {P25} radio systems in the
  presence of fireground noise: Test plan and results,'' National
  Telecommunications and Information Administration, Tech. Rep. TR-13-495, May
  2013. [Online]. Available:
  \url{https://its.ntia.gov/publications/download/TR-13-495.pdf}
\BIBentrySTDinterwordspacing

\bibitem{ntia2012:amr-intelligibility}
\BIBentryALTinterwordspacing
D.~J. Atkinson, S.~D. Voran, and A.~A. Catellier, ``Intelligibility of the
  adaptive multi-rate speech coder in emergency-response environments,''
  National Telecommunications and Information Administration, Tech. Rep.
  TR-13-493, Dec. 2012. [Online]. Available:
  \url{https://its.ntia.gov/publications/download/13-493.pdf}
\BIBentrySTDinterwordspacing

\bibitem{voran2015codecs}
S.~D. Voran and A.~A. Catellier, ``Speech codec intelligibility testing in
  support of mission-critical voice applications for lte,'' National
  Telecommunications and Information Administration (NTIA), NTIA Report 15-520,
  Sep. 2015.

\bibitem{ntia2015:codecs-lte}
\BIBentryALTinterwordspacing
------, ``Speech codec intelligibility testing in support of mission-critical
  voice applications for {LTE},'' National Telecommunications and Information
  Administration, Tech. Rep. TR-15-520, Sep. 2015. [Online]. Available:
  \url{https://its.ntia.gov/publications/download/TR-15-520.pdf}
\BIBentrySTDinterwordspacing

\bibitem{ntia2017:codecs}
\BIBentryALTinterwordspacing
------, ``Intelligibility robustness of five speech codec modes in
  frame-erasure and background-noise environments,'' National
  Telecommunications and Information Administration, Tech. Rep. TR-18-529, Dec.
  2017. [Online]. Available:
  \url{https://its.ntia.gov/publications/download/TR-18-529.pdf}
\BIBentrySTDinterwordspacing

\bibitem{ntia2009:speaker-identification}
\BIBentryALTinterwordspacing
A.~A. Catellier and S.~D. Voran, ``Relationships between intelligibility,
  speaker identification, and the detection of dramatized urgency,'' National
  Telecommunications and Information Administration, Tech. Rep. TR-09-459, Mar.
  2009. [Online]. Available:
  \url{https://its.ntia.gov/publications/download/TR-09-459.pdf}
\BIBentrySTDinterwordspacing

\bibitem{voran2008listener}
S.~Voran, ``Listener detection of talker stress in low-rate coded speech,'' in
  \emph{2008 IEEE International Conference on Acoustics, Speech and Signal
  Processing}.\hskip 1em plus 0.5em minus 0.4em\relax IEEE, 2008, pp.
  4813--4816.

\bibitem{catellier2008speaker}
A.~Catellier and S.~Voran, ``Speaker identification in low-rate coded speech,''
  in \emph{Proc. 7th International Measurement of Audio and Video Quality in
  Networks Conference}, 2008.

\bibitem{mauch2013audio}
M.~Mauch and S.~Ewert, ``{The Audio Degradation Toolbox and its Application to
  Robustness Evaluation},'' in \emph{Proceedings of the 13th International
  Society for Music Information Retrieval}, Nov. 2013.

\end{thebibliography}


\end{document}